\documentclass[sigconf]{acmart}

 \usepackage{amsmath}
  \usepackage{multirow}
\usepackage{epstopdf}
  \usepackage{graphicx}
\usepackage{appendix}
\usepackage[normalem]{ulem}
\useunder{\uline}{\ul}{}
 \usepackage{enumitem}
 \useunder{\uline}{\ul}{}
\AtBeginDocument{%
  \providecommand\BibTeX{{%
    \normalfont B\kern-0.5em{\scshape i\kern-0.25em b}\kern-0.8em\TeX}}}
\newcommand{\para}[1]{\noindent \textbf{#1}}
\usepackage{listings}
\usepackage{color}

\definecolor{dkgreen}{rgb}{0,0.6,0}
\definecolor{gray}{rgb}{0.5,0.5,0.5}
\definecolor{mauve}{rgb}{0.58,0,0.82}

\lstdefinestyle{Python}{
    language        =   Python, %
    basicstyle      =   \ttfamily,
    numberstyle     =   \ttfamily,
    keywordstyle    =   \color{blue},
    keywordstyle    =   [2] \color{teal},
    stringstyle     =   \color{magenta},
    commentstyle    =   \color{red}\ttfamily,
    breaklines      =   true,   %
    columns         =   fixed,  %
    basewidth       =   0.5em,
}

\usepackage{subfigure}

\usepackage{hyperref}
\usepackage{url}

\copyrightyear{2023} 
\acmYear{2023}
\setcopyright{rightsretained}
\acmConference[WWW '23]{Proceedings of the ACM Web Conference
2023}{May 1--5, 2023}{Austin, TX, USA}
\acmBooktitle{Proceedings of the ACM Web Conference 2023 (WWW '23),
May 1--5, 2023, Austin, TX, USA}
\acmDOI{10.1145/3543507.3583278}
\acmISBN{978-1-4503-9416-1/23/04}

\usepackage{etoolbox}
\makeatletter
\patchcmd{\maketitle}{\@copyrightpermission}{
   \begin{minipage}{0.3\columnwidth}
     \href{http://creativecommons.org/licenses/by/4.0/}{\includegraphics[width=0.90\textwidth]{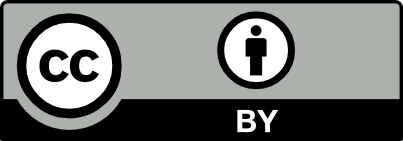}}
   \end{minipage}\hfill
   \begin{minipage}{0.7\columnwidth}
     \href{http://creativecommons.org/licenses/by/4.0/}{This work is licensed under a Creative Commons Attribution International 4.0 License.}
   \end{minipage}
  
   \vspace{5pt}
}{}{}

\makeatother

\author{Guanyu Lin$^{1}$, Chen Gao$^{1\dagger}$, Yu Zheng$^{1}$, Jianxin Chang$^{3}$, Yanan Niu$^{3}$, Yang Song$^{3}$, \\Zhiheng Li$^{2}$, Depeng Jin$^{1}$, Yong Li$^{1}$}
\affiliation{
 \institution{$^1$Department of Electronic Engineering, Beijing National Research Center for Information Science and Technology, Tsinghua University}
  \institution{$^2$International Graduate School, Tsinghua University}
 \institution{$^3$Beijing Kuaishou Technology Co., Ltd.}
 \country{}
}
\thanks{$\dagger$Chen Gao is the corresponding author (chgao96@gmail.com).}
\renewcommand{\authors}{Lin, et al} %
\begin{document}
\fancyhead{}

\title{Dual-interest Factorization-heads Attention for Sequential Recommendation}

\begin{abstract}
Accurate user interest modeling is vital for recommendation scenarios. One of the effective solutions is the sequential recommendation that relies on click behaviors, but this is not elegant in the video feed recommendation where users are passive in receiving the streaming contents and return skip or no-skip behaviors. Here skip and no-skip behaviors can be treated as negative and positive feedback, respectively. With the mixture of positive and negative feedback, it is challenging to capture the transition pattern of behavioral sequence. To do so, FeedRec has exploited a shared vanilla Transformer, which may be inelegant because head interaction of multi-heads attention does not consider different types of feedback. In this paper, we propose \textbf{D}ual-interest \textbf{F}actorization-heads \textbf{A}ttention for Sequential \textbf{R}ecommendation (short for DFAR) consisting of feedback-aware encoding layer, dual-interest disentangling layer and prediction layer. In the feedback-aware encoding layer, we first suppose each head of multi-heads attention can capture specific feedback relations. Then we further propose factorization-heads attention which can mask specific head interaction and inject feedback information so as to factorize the relation between different types of feedback. Additionally, we propose a dual-interest disentangling layer to decouple positive and negative interests before performing disentanglement on their representations. Finally, we evolve the positive and negative interests by corresponding towers whose outputs are contrastive by BPR loss. Experiments on two real-world datasets show the superiority of our proposed method against state-of-the-art baselines. Further ablation study and visualization also sustain its effectiveness. We release the source code here: {\url{https://github.com/tsinghua-fib-lab/WWW2023-DFAR}}.
\end{abstract}




\maketitle

\vspace{-0.2cm}
\section{Introduction}
\begin{figure}[t!]
		\centering
		\begin{tabular}{c}
		    	\includegraphics[width=\columnwidth]{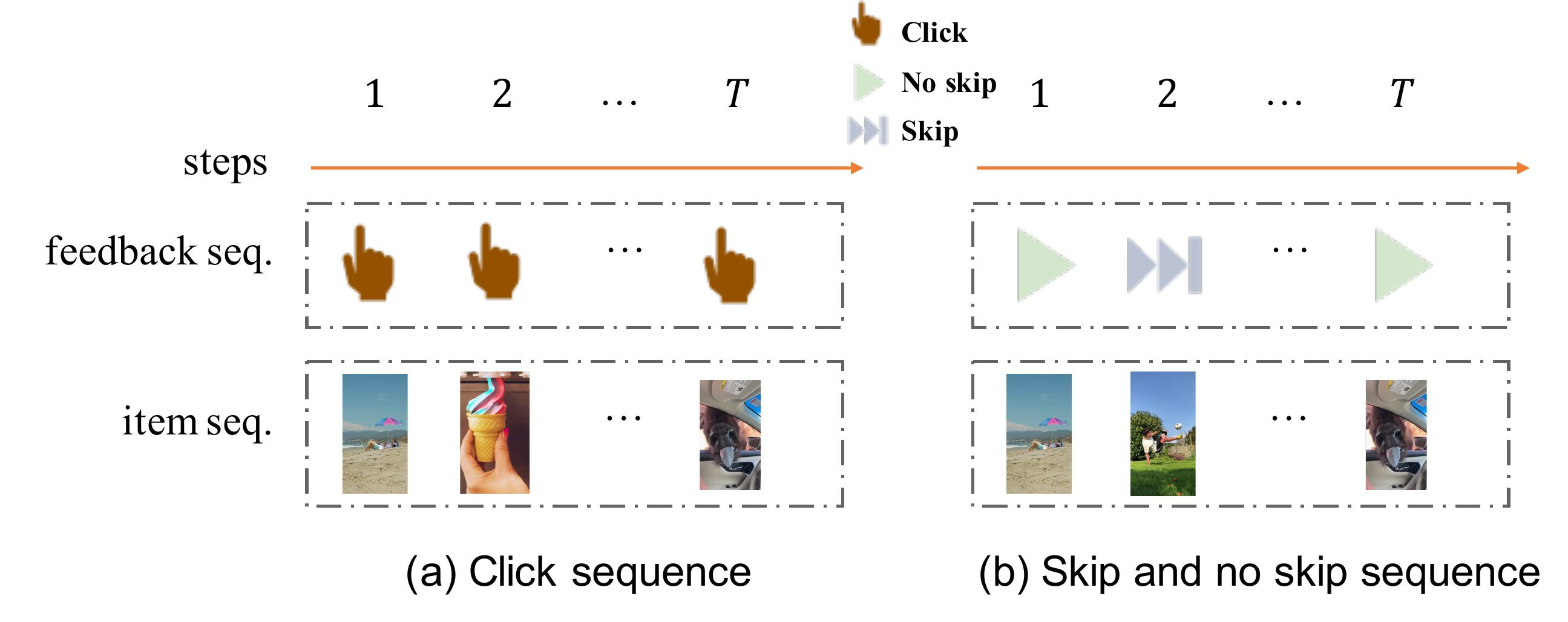}
		\end{tabular}
%  \vspace{-0.5cm}
				\setlength{\abovecaptionskip}{-0.3cm}
		\setlength{\belowcaptionskip}{-0.5cm}
	\caption{Illustration of click-based sequential recommendation and our dual-interest sequential recommendation which is hybrid with positive and negative feedback.}	\label{fig:video}
%  \vspace{-0.4cm}

\end{figure}

\vspace{-0.1cm}
Online sequential recommendation~\cite{SRs} has achieved great success for its time-aware personalized modeling and has been widely applied in Web platforms, including micro-video, news, e-commerce, etc. Especially in today's video feed recommendation, users are attracted immensely by video streaming which can be treated as a sequence of items. 
Formally speaking, the sequential recommendation is defined as predicting the next interacted item by calculating the matching probability between historical items and the target item.
As shown in Figure~\ref{fig:video} (a), existing sequential recommendation models often exploit click behaviors of users to infer their dynamic interests~\cite{DIN, wang2020fine, DIEN, GRU4REC, SASRec}, the optimization of which samples un-clicked items as negative feedback. 
However, such an approach only inputs positive items into the sequential model, and negative items are sampled as target items, ignoring the transition pattern between historical positive and negative items.

In the video feed recommendation where a single item is exposed each time, users either skip or do not skip the recommended items, as illustrated in Figure~\ref{fig:video} (b). Skip can be treated as a kind of negative feedback which means users don't want to receive certain items, while no-skip can be treated as a kind of positive feedback. That is to say, users are passive in receiving the recommended items without providing active click behaviors~\cite{li2016context, okura2017embedding, guo2019streaming} in such video feed recommendations. However, the existing click-based sequential recommendation does not consider the transition pattern between positive and negative items. Indeed, there are two key challenges when modeling positive and negative feedback in one sequence.
\vspace{-0.05cm}
\begin{itemize}[leftmargin=*] 
 \item \textbf{Complex transition between positive and negative feedback.} The transition pattern among interacted items has become far more complex due to negative feedback. A user may provide negative feedback only because she has consumed a very similar item before, which makes accurate modeling of transition essential and challenging.

\item \textbf{Mixed interest in one behavioral sequence.}
The negative feedback in the behavioral sequence brings significant challenges to interest learning. The traditional methods of sequential recommendation always conduct a pooling operation on user sequence to obtain the users' current interest, which will fail when the sequence is hybrid with positive and negative signals.
\end{itemize}
\vspace{-0.05cm}

To address the above challenges, in this work, we propose a model named \textbf{D}ual-interest \textbf{F}actorization-heads \textbf{A}ttention for Sequential \textbf{R}ecommendation (short for DFAR),
further extracting the transition pattern and pair-wise relation between positive and negative interests. 
To address the first challenge, in the feedback-aware encoding layer, we assume each head of multi-head attention~\cite{vaswani2017attention} tends to capture specific relations of certain feedback~\cite{voita-etal-2019-analyzing}.
As different heads of multi-head attention~\cite{vaswani2017attention} are independent, it may fail to capture the transition pattern between different feedback when positive feedback and negative feedback are indeed not independent of each other.
Thus we exploit talking-heads attention~\cite{shazeer2020talking} to implicitly extract the transition pattern between positive and negative historical items. However, talking-heads attention may mix different heads too much without sufficient prior knowledge. To explicitly extract the transition pattern between positive and negative historical items, we further propose feedback-aware factorization-heads attention which can even incorporate the feedback information into the head interaction. 
To address the second challenge, we propose a dual-interest disentangling layer and prediction layer, respectively, to disentangle and extract the pair-wise relation between positive and negative interests.
Specifically, we first mask and encode the sequence hybrid with positive feedback and negative feedback into two single interest representations before performing disentanglement on them to repel the dissimilar interests. Then we perform a prediction of each interest with the corresponding positive or negative tower and apply contrastive loss on them to extract their pair-wise relation.

In general, we make the following contributions in this work.
\vspace{-0.05cm}
\begin{itemize}[leftmargin=*] 
    \item  We have taken the pioneering step of fully considering the modeling of negative feedback, along with its impact on transition patterns, to enhance sequential recommendation.
    \item  We propose a feedback-aware encoding layer to capture the transition pattern, dual-interest disentangling layer and prediction layer to perform disentanglement and capture the pair-wise relation between positive and negative historical items.
    \item We conduct experiments on one benchmark dataset and one collected industrial dataset, where the results show the superiority of our proposed method. A further ablation study also sustains the effectiveness of our three components.
\end{itemize}
\vspace{-0.05cm}
\vspace{-0.1cm}
\section{Problem Formulation}
\vspace{-0.05cm}
\para{\textbf{Click-based Sequential Recommendation.}}
Given item sequence $\mathcal{I}_{u} = (i_{1}, i_{2}, \ldots, i_{t})$ with only positive feedback, the goal of traditional click-based sequential recommendation is accurately predicting the probability that \textbf{given user} $u$ will click the target item \textit{i.e.}, $i_{t+1}$. 
The traditional click-based sequential recommendation can be formulated as follows.

\noindent \textit{\textbf{Input}}: Item sequence $\mathcal{I}_{u} = (i_{1}, i_{2}, \ldots, i_{t})$ with only positive feedback for a \textbf{given user $u$}.
    
    \noindent \textit{\textbf{Output}}: The predicted score that the \textbf{given user $u$} will click the target item $i_{t+1}$.

\para{\textbf{Dual-interest Sequential Recommendation.}}
Given item sequence $\mathcal{I}_{u} = (i_{1}, i_{2}, \ldots, i_{t})$ with both positive and negative feedback, the dual-interest sequential recommendation aims to better predict the probability that \textbf{given user} $u$ will skip or not skip the target item \textit{i.e.}, $i_{t+1}$. 
The dual-interest sequential recommendation with both positive and negative feedback can be formulated as follows.

    \noindent \textit{\textbf{Input}}: Item sequence $\mathcal{I}_{u} = (i_{1}, i_{2}, \ldots, i_{t})$ with positive and negative feedbacks for a \textbf{given user $u$}.
    
    \noindent \textit{\textbf{Output}}: The predicted score that the \textbf{given user $u$} will skip or do not skip the target item $i_{t+1}$.
\section{Methodology}
\begin{figure*}[t!]
\centering
\renewcommand\arraystretch{0.01}

\begin{tabular}{c}
     	\includegraphics[width=.9\linewidth]{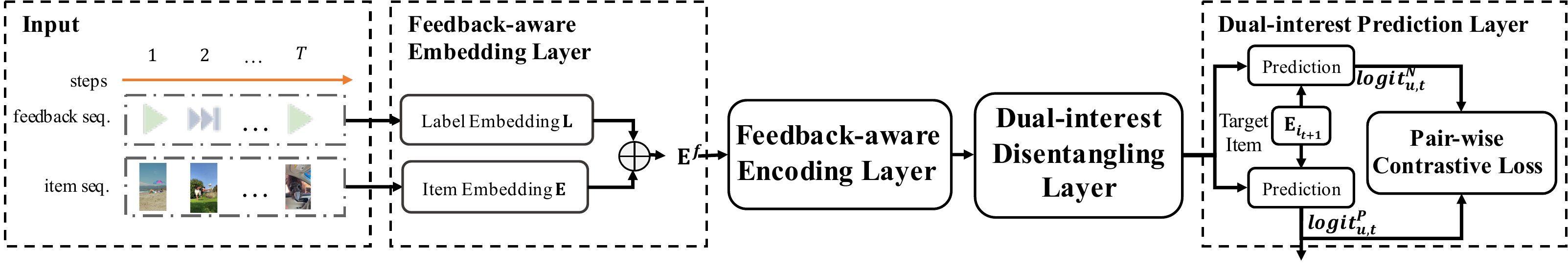} 
	\\
     	\includegraphics[width=.9\linewidth]{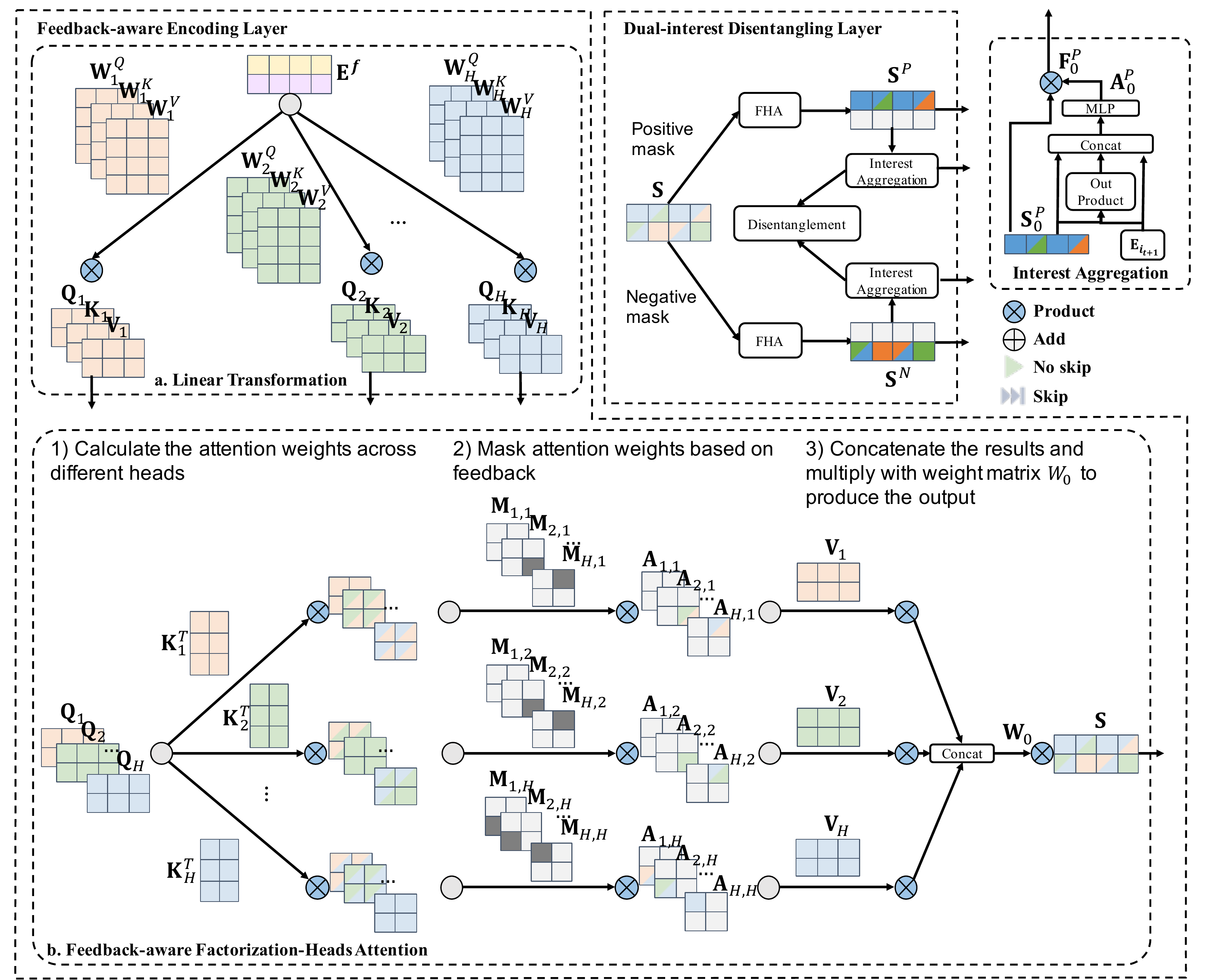} 
	
\end{tabular}
	\caption{Illustration of DFAR. ($\romannumeral1$) \textbf{Feedback-aware Encoding Layer} is linked after the Feedback-aware Embedding Layer where each historical item is injected with a label embedding according to the corresponding feedback; It consists of linear transformation and feedback-aware factorization-heads attention. In the linear transformation, input embeddings are transformed into query, key and value matrices. In feedback-aware factorization-heads attention, the transition relation between different items is factorized into different heads which are masked according to the positive or negative feedback. ($\romannumeral2$) \textbf{Dual-interest Disentangling Layer} decouples positive and negative interests and performs disentanglement to repel the dissimilar representations of different feedback; ($\romannumeral3$) \textbf{Dual-interest Prediction Layer} evolves positive and negative interests with corresponding towers and perform BPR loss to capture the pair-wise relation.}	\label{fig:architecture}
\end{figure*}

Our model captures the relation between positive feedback and negative feedback at the transition level and interest level of sequential recommendation, respectively, by the proposed Feedback-aware Encoding Layer, Dual-interest Disentangling Layer and Prediction Layer, as shown in Figure~\ref{fig:architecture}. 

\begin{itemize}[leftmargin=*]
\item \textbf{Feedback-aware Encoding Layer}.
We build item embeddings by item IDs and label embeddings by item feedback and further propose feedback-aware factorization-heads attention to capture the transition pattern between different feedback.

\item \textbf{Dual-interest Disentangling Layer}. 
 We mask the sequence hybrid with both positive and negative feedback into two sequences with solely positive or negative feedback. After encoding two split sequences with independent factorization-heads attention to extract the positive and negative interests, we then disentangle them to repel the dissimilar interests.

 \item \textbf{Dual-interest Prediction Layer}.
 We further extract the positive and negative interests with independent towers and then perform contrastive loss on them to extract the pair-wise relation.
\end{itemize}
\subsection{Feedback-aware Encoding Layer}
In the feedback-aware encoding layer, we first inject each historical item embedding with corresponding feedback embeddings to incorporate the feedback information into each historical item embedding. Then we further propose talking-heads attention and feedback-aware factorization-heads attention to capture the transition pattern between positive and negative historical items.
\subsubsection{\textbf{Feedback-aware Embedding Layer}}
To fully distinguish positive and negative feedback, we build a label embedding matrix $\textbf{L} \in \mathbb{R}^{2 \times D}$, besides the 
item embedding matrix $\textbf{E} \in \mathbb{R}^{ m \times D}$. Here $m$ denotes the number of items, and $D$ is the dimensionality for the hidden state. 
Then we inject the feedback information into the item embedding and obtain the feedback-aware input embeddings as the model input. 
Therefore, given item sequence $\mathcal{I}_{u} = (i_{1}, i_{2}, \ldots, i_{t})$, we can obtain the feedback-aware item embeddings $\mathbf{E}^f \in \mathbb{R}^{T \times D}$ as:
\begin{equation}\label{eqn:emb}
\begin{aligned}
\mathbf{E}^f &= [\textbf{E}_{i_{1}} , \textbf{E}_{i_{2}} , \ldots, \textbf{E}_{i_{t}} ] + [\textbf{L}_{y_{u, i_{1}}} , \textbf{L}_{y_{u,i_{2}}} , \ldots, \textbf{L}_{y_{u, i_{t}}} ],
\end{aligned}
\end{equation}
where  $\{y_{u, i_{1}}, y_{u,i_{2}}, \cdots, y_{u, i_{t}}\}$ are feedback of items $\{i_{1}, i_{2}, \cdots, i_{t}\}$. Here $y_{u, i_{1}} = 1$ if $i_{1}$ is the no-skip item, and $y_{u, i_{1}} = 0$ if $i_{1}$ is the skip item.
Note that if the sequence length is less than $t$, we can pad $\mathbf{E}^f$ with zero embedding~\cite{SASRec}. 

\subsubsection{\textbf{Talking-Heads Attention}}~\label{sec:talking}
After obtaining the input embeddings for positive and negative historical items, we then capture the transition pattern between them. The existing work,  FeedRec~\cite{wu2022feedrec},
exploits vanilla Transformer to roughly capture this transition pattern, of which multi-head attention~\cite{SASRec} is the essential part, having the following equation:
\begin{equation}
\begin{array}{l}
\textbf{S} = \operatorname{MHA}(\mathbf{Q}, \mathbf{K}, \mathbf{V})=\left[\mathbf{A}_{1}^{MHA} \mathbf{V}_1, \ldots, \mathbf{A}_{H}^{MHA} \mathbf{V}_H\right]  \mathbf{W}_0,
\end{array}
\end{equation}
 \begin{equation}
 \mathbf{A}^{MHA}_h =\operatorname{softmax}\left(  \frac{ \mathbf{Q}_h {\mathbf{K}_h}^T}{\sqrt{d}}\right), 
\end{equation}
 \begin{equation}
 \mathbf{Q}_h = \mathbf{Q} \mathbf{W}_h^Q, \mathbf{K}_h = \mathbf{K} \mathbf{W}_h^K, \mathbf{V}_h = \mathbf{V} \mathbf{W}_h^V,
\end{equation}
where $ h \in \{1, 2, \cdots, H\}$ is the number of heads. $\mathbf{W}_0 \in \mathbb{R}^{HD \times D}$ and $\mathbf{W}_h^Q$, $\mathbf{W}_h^K$,  $\mathbf{W}_h^V \in \mathbb{R}^{D \times D}$ are parameters to be learned. $\operatorname{MHA}$ means multi-heads attention~\cite{vaswani2017attention}.
However, different heads of multi-head attention are independent of each other, sharing no information across heads. If assuming different heads capture specific relations between different feedback, then this means there is no information sharing across different feedback. Thus we first propose talking-heads attention~\cite{shazeer2020talking} to address this issue as below. 
\begin{equation}
\begin{array}{l}
\textbf{S} = \operatorname{THA}(\mathbf{Q}, \mathbf{K}, \mathbf{V})=\left[\mathbf{A}_{1}^{THA} \mathbf{V}_1, \ldots, \mathbf{A}_{H}^{THA} \mathbf{V}_H\right] \mathbf{W}_0,
\end{array} 
\end{equation}
\begin{equation}\label{eq:before_softmax}
\begin{aligned}
   \left[\begin{array}{c}
	\mathbf{A}_1 \\
	\mathbf{A}_2\\
	\vdots\\
	\mathbf{A}_{H'}
	\end{array}\right] = \mathbf{W}_{THA}
    \left[\begin{array}{c}
	\frac{ \mathbf{Q}_1 {\mathbf{K}_1}^T}{\sqrt{d}} \\
		\frac{ \mathbf{Q}_2 {\mathbf{K}_2}^T}{\sqrt{d}}\\
	\vdots\\
		\frac{ \mathbf{Q}_H {\mathbf{K}_H}^T}{\sqrt{d}}
	\end{array}\right], 
	\end{aligned} 
\end{equation}
\begin{equation}\label{eq:after_softmax}
    	\begin{aligned}
   \left[\begin{array}{c}
	\mathbf{A}_1^{THA} \\
	\mathbf{A}_2^{THA}\\
	\vdots\\
	\mathbf{A}_{H}^{THA}
	\end{array}\right] = \mathbf{W}_{THA}^{S}
    \left[\begin{array}{c}
	\operatorname{softmax}\left( \mathbf{A}_1\right)  \\
	\operatorname{softmax}\left( \mathbf{A}_2\right)\\
	\vdots\\
	\operatorname{softmax}\left( \mathbf{A}_{H'}\right)
	\end{array}\right], 
	\end{aligned}
\end{equation}
where $\mathbf{W}_{THA} \in \mathbb{R}^{H' \times H}$, $\mathbf{W}^{S}_{THA} \in \mathbb{R}^{H \times H'}$ and $\mathbf{W}_0 \in \mathbb{R}^{HD \times D}$  are parameters to be learned. Here $ \operatorname{THA}$ refers to talking-heads attention.
 However, the interaction between different heads in talking-heads attention is implicit, which may confuse the task for each head and result in overfitting. Not to mention, the two additional linear transformations (i.e. Eq.\eqref{eq:before_softmax} and Eq.\eqref{eq:after_softmax}) of talking-heads attention will increase the computation cost.

\subsubsection{\textbf{Feedback-aware Factorization-heads Attention}} 

\begin{figure}[t!]
\centering

\begin{tabular}{c}
     	\includegraphics[width=.9\columnwidth]{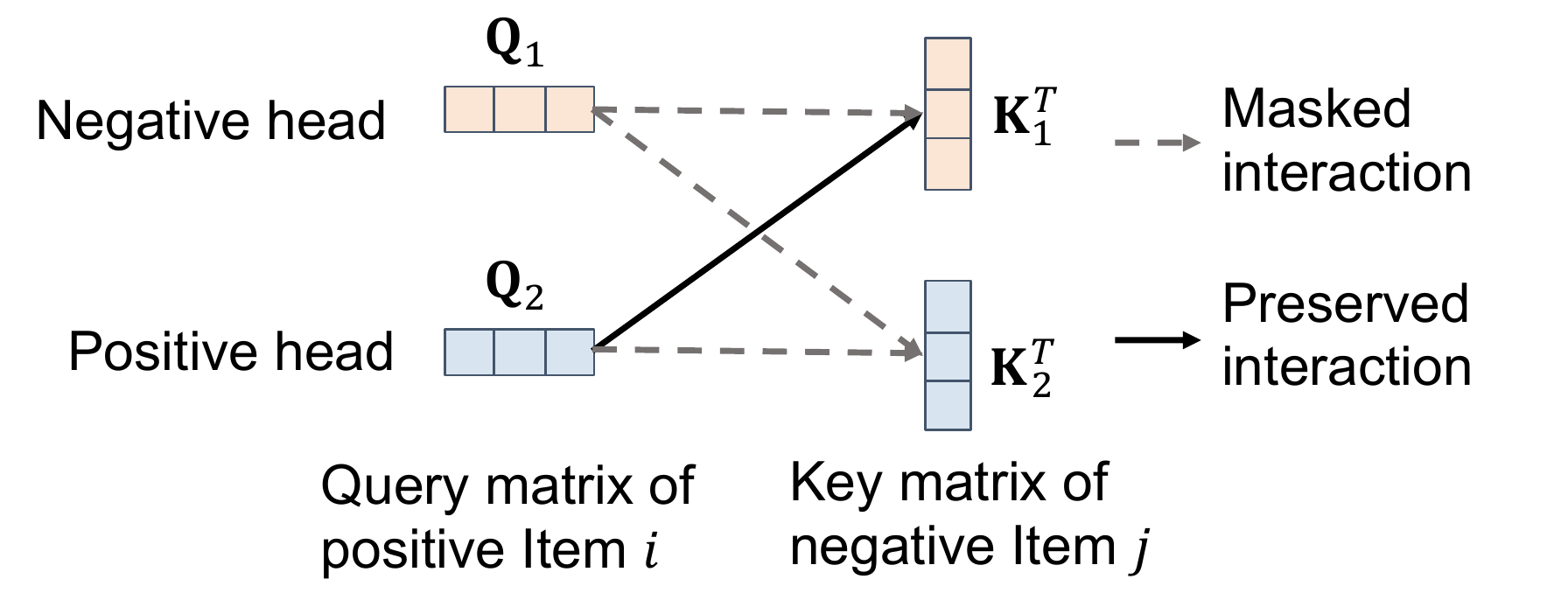}

\end{tabular}
		
		\setlength{\abovecaptionskip}{-0.2cm}
        \setlength{\belowcaptionskip}{-0.2cm}	\caption{Illustration of label mask $\mathbf{M}_{h_1, h_2}$ on head interaction. Here we show the comprehensible case with two heads, where the first half of heads, i.e. head 1, represents negative head and second half of heads, i.e. head 2, represents positive head.}	\label{fig:head_mask}
\end{figure}

In this part, we factorize the interaction between positive and negative feedback. Traditional multi-heads attention assigns similar items with higher attention weights. However, in our problem with both positive and negative feedback, two similar items may have different attention weights due to the feedback they have. For example, an NBA fan skips the recommended video about basketball when he/she has watched a lot of basketball videos. But he/she engages in the video about basketball when he/she only has watched a few videos about basketball. 
In the first case we should repel the representations between historical basketball videos and target basketball videos, while in the second case we should attract them. That is to say, it is necessary to inject the user's feedback into the transition pattern between different feedback.
Here we suppose different heads can represent different transition patterns for different feedback~\cite{voita-etal-2019-analyzing}.
To explicitly factorize interaction across different heads, we further propose factorization-heads attention as:
\begin{equation}\label{eq:fha}
\begin{array}{l}
\textbf{S} = \operatorname{FHA}(\mathbf{Q}, \mathbf{K}, \mathbf{V})=\left[\mathbf{A}^{FHA}_{1, 1} \mathbf{V}_{1}, \ldots, \mathbf{A}^{FHA}_{{H}, {H}} \mathbf{V}_{{H}}\right] \mathbf{W}_0,
\end{array}
\end{equation}
 \begin{equation}
 \mathbf{A}^{FHA}_{h_1, h_2} =\operatorname{softmax}\left(  \frac{ \mathbf{Q}_{h_1} {\mathbf{K}_{h_2}}^T}{\sqrt{d}}\right), 
\end{equation}
where ${h_1, h_2} \in \{1, 2, \cdots, H\}$. $\mathbf{W}_0 \in \mathbb{R}^{HD \times D}$ are parameters to be learned. Here $\operatorname{FHA}$ is our proposed factorization-heads attention. The factorization-heads attention can represent $H \times H$ relations by $H$ heads. That is to say, our factorization-heads attention can reduce $\sqrt{H}$ times parameters if we want to represent $H$ head interaction relations like talking-heads attention or multi-heads attention.
Besides, to further inject the prior feedback knowledge into the factorization-heads attention, we propose feedback-aware factorization-heads attention with a label mask  $\mathbf{M}_{h_1, h_2} \in \{0, 1\}^{t \times t}$ as:
\begin{equation}
\begin{array}{l}
\textbf{S} = \operatorname{FFHA}(\mathbf{Q}, \mathbf{K}, \mathbf{V})=\left[\mathbf{A}^{FFHA}_{1, 1} \mathbf{V}_{1}, \ldots, \mathbf{A}^{FFHA}_{{H}, {H}} \mathbf{V}_{{H}}\right] \mathbf{W}_0,
\end{array}
\end{equation}
 \begin{equation}\label{eq:attention_heads}
 \mathbf{A}^{FFHA}_{h_1, h_2} =\operatorname{softmax}\left( \mathbf{M}_{h_1, h_2} \frac{ \mathbf{Q}_{h_1} {\mathbf{K}_{h_2}}^T}{\sqrt{d}}\right), 
\end{equation}
where $\mathbf{M}_{{h_1, h_2}, i,j} = 1$, if $h_1 \in \{  \frac{y_{u,i}{H}}{2} + 1, \frac{y_{u,i}{H}}{2} + 2, \cdots,  \frac{(y_{u,i} + 1){H}}{2} \}, h_2 \in \{\frac{y_{u,j}{H}}{2} +1,  \frac{y_{u,j}{H}}{2} + 2, \cdots,  \frac{(y_{u,j} + 1){H}}{2} \}, i \in \{1, 2, \cdots, t\}, j \in \{1, 2, \cdots, t\}$ and $\mathbf{M}_{{h_1, h_2},i,j} = 0$, otherwise. 
Here the first half of heads w.r.t.  $\{1,2,\cdots,\frac{H}{2}\}$ represent negative heads and second half of heads w.r.t. $\{\frac{H}{2}+1, \frac{H}{2}+2,\cdots,H\}$ represent positive heads. For example, as shown in Figure~\ref{fig:head_mask}, if item $i$ is positive and item $j$ is negative (i.e., $y_{u,i} = 1$ and $y_{u,j} = 0$), $h_1$ in positive half and $h_2$ in negative half will be preserved, i.e., 
$\mathbf{M}_{{2, 1}, i,j} = 1$, and $\mathbf{M}_{{1, 1}, i,j}, \mathbf{M}_{{1, 2}, i,j}, \mathbf{M}_{{2, 2}, i,j} = 0$.

Besides, $\operatorname{FFHA}$ is our proposed feedback-aware factorization-heads attention.
Apart from the advantage of explicit interaction between different heads, unlike talking-heads attention, our factorization-heads attention also improves the multi-heads attention without high computation cost. We feed the input embedding into the feedback-aware factorization attention module as:
\begin{equation}
\begin{array}{l}
\mathbf{S} = \operatorname{FFHA}(\mathbf{E}^f, \mathbf{E}^f, \mathbf{E}^f),
\end{array}
\end{equation}
where $\mathbf{S}$ are the obtained feedback-aware sequential representations. We put the pseudocode of FHA at Appendix~\ref{sec:pseudocode} and compare its complexity with MHA and THA at Appendix~\ref{sec:compare}.

\subsection{Dual-interest Disentangling Layer}
Though feedback-aware factorization-heads attention has factorized the transition relation between positive feedback and negative feedback, their interest-level relations require further extracting.
In this part, we decouple the positive and negative interests and then perform disentanglement on them to repel the dissimilar interests. 

\subsubsection{\textbf{Dual-interest Decoupling Attention}}
After capturing the transition pattern between positive feedback and negative feedback, we then filter out each feedback by a corresponding feedback mask as follows,
\begin{equation}\label{eqn:masked}
\begin{aligned}
\mathbf{S}^P &= [\textbf{S}_{i_{1}} , \textbf{S}_{i_{2}} , \ldots, \textbf{S}_{i_{t}} ] * [{y_{u, i_{1}}} , {y_{u,i_{2}}} , \ldots, {y_{u, i_{t}}} ],\\
\mathbf{S}^N &= [\textbf{S}_{i_{1}} , \textbf{S}_{i_{2}} , \ldots, \textbf{S}_{i_{t}} ] * (1 - [{y_{u, i_{1}}} , {y_{u,i_{2}}} , \ldots, {y_{u, i_{t}}} ]),
\end{aligned}
\end{equation}
which are then fed into the corresponding factorization-heads attention modules to enhance the transition pattern learning for each feedback as:
\begin{equation}
\begin{array}{l}
\mathbf{S}^P = \operatorname{FHA}(\mathbf{S}^P, \mathbf{S}^P, \mathbf{S}^P), \mathbf{S}^N = \operatorname{FHA}(\mathbf{S}^N, \mathbf{S}^N, \mathbf{S}^N),

\end{array}
\end{equation}
where $\mathbf{S}^P$ (or $\mathbf{S}^N$) are the single-feedback sequential representations for positive feedback (or negative feedback). 
In the subsequent section, we will exploit these filtered representations to further extract the interest-level relations.

\subsubsection{\textbf{Dual-interest Aggregation and Disentanglement}}
The positive and negative interests of a given user should be distinguished from each other. Hence we aim to repel the positive and negative representations of corresponding interests. 
Specifically, we assume the target item is possibly either positive or negative. Then we assign the target item with positive and negative label embeddings, respectively, in positive and negative assumed cases. To calculate the attention scores of positive and negative historical items, we fuse them with the target item in assumed positive and negative cases as below.
\begin{equation}\label{eq:similarScore_A}
	\mathbf{A}^P=\text{MLP}\left( (\textbf{E}_{i_{t+1}} + \textbf{L}_1) \| \mathbf{S}^{P} \right) ,
	\mathbf{A}^N=\text{MLP}\left( (\textbf{E}_{i_{t+1}} + \textbf{L}_0) \| \mathbf{S}^{N} \right) ,
\end{equation}
where $\mathbf{A}^P$ and $\mathbf{A}^N \in \mathbb{R}^{t \times D}$ are the positive and negative attention scores. $\text{MLP}$ is the multi-layer perceptron. Here $\textbf{L}_1$ and $\textbf{L}_0$ are the label embeddings for positive and negative feedback, respectively.
With the calculated attention scores by \eqref{eq:similarScore_A}, we can then obtain the single-feedback aggregated representations for positive and negative items, respectively, as,
\begin{equation}\label{eq:similarItem} \small
	\mathbf{F}^{P}=\textbf{softmax}\left( \mathbf{A}^{P}\right)\mathbf{S}^{P}   ,
	\mathbf{F}^{N}=\textbf{softmax}\left( \mathbf{A}^{N}\right)\mathbf{S}^{N},
\end{equation}
\begin{equation}\label{eq:single_aggre} 
\begin{aligned}
 \mathbf{f}^{P} = \sum_{j=1}^{t} \mathbf{F}^{P}_j, \mathbf{f}^{N} = \sum_{j=1}^{t} \mathbf{F}^{N}_j,
 \end{aligned} 
\end{equation}
which are then further disentangled with cosine distance as:
\begin{equation}
\mathcal{L}^D=\frac{\mathbf{f}^P \cdot \mathbf{f}^N}{\left\|\mathbf{f}^P\right\| \times\left\|\mathbf{f}^N\right\|}.
\end{equation}
where $\|\cdot\|$ is the L2-norm. By this disentangling loss, we can repel the aggregated positive and negative representations so as to capture the dissimilar characteristics between them.

\subsection{\textbf{Dual-interest Prediction Layer}}

In this section, we predict the next item of different interests by positive and negative towers. Finally, we further perform contrastive loss on the outputs of positive and negative towers so as to extract the pair-wise relation between them.
\subsubsection{\textbf{Dual-interest Prediction Towers}}
To extract the positive and negative interests, we fuse the feedback-aware sequential representations, single-feedback sequential representations, and single-feedback aggregated representations into the corresponding positive or negative prediction tower.
Before feeding different representations into the final prediction towers, we first aggregate part of them by the sum pooling as:
$$ 
\begin{aligned}
\mathbf{s}= \sum_{j=1}^{t} \mathbf{S}_j, \mathbf{s}^{P}= \sum_{j=1}^{t} \mathbf{S}^{P}_j, \mathbf{s}^{N}= \sum_{j=1}^{t} \mathbf{S}^{N}_j, 
 \end{aligned} 
$$
which are then finally fed into the positive and negative prediction towers as:
\begin{equation}
	\textit{logit}^P_{u, t}=\textbf{MLP}\left(\mathbf{s} \| \mathbf{s}^{P}\|\mathbf{f}^{P} \|  (\textbf{E}_{i_{t+1}} + \textbf{L}_1) \right),
\end{equation}
\begin{equation}
	\textit{logit}^N_{u, t}=\textbf{MLP}\left(\mathbf{s} \| \mathbf{s}^{N}\|\mathbf{f}^{N} \|  (\textbf{E}_{i_{t+1}} + \textbf{L}_0) \right).
\end{equation}
where $\textit{logit}^P_{u, t}$ and $\textit{logit}^N_{u, t}$ are positive and negative predicted logits for user $u$ on time step $t$, aiming to capture the positive and negative interests, respectively. Here $\mathbf{f}^{P}$ and $\mathbf{f}^{N}$ are pooled at Eq.\eqref{eq:single_aggre}.

\subsubsection{\textbf{Pair-wise Contrastive Loss}}
When the target item is positive, the prediction logit of the positive tower will be greater than that of the negative tower, and vice versa. After obtaining the positive and negative prediction logits, we then perform BPR loss~\cite{BPR} on them as: 

\begin{equation}\label{eq:contrastive_loss}
\mathcal{L}^{BPR} = \left\{
\begin{aligned}
-\log(\sigma(\textit{logit}^P_{u, t} - \textit{logit}^N_{u, t}))& , & {y}_{u, t} = 1, \\
-\log(\sigma(\textit{logit}^N_{u, t} - \textit{logit}^P_{u, t})) & , & {y}_{u, t} = 0.
\end{aligned}
\right.
\end{equation}
where $\sigma$ denotes the sigmoid function. With this BPR loss, we can extract the pair-wise relations between positive and negative logits.

\subsection{Joint Optimization}

Though we have positive and negative towers, in the optimization step, we only need to optimize the next item prediction loss with the positive tower as:
\begin{equation}\label{eq:loss}
\mathcal{L}=-\frac{1}{|\mathcal{R}|} \sum_{(u, i_t) \in \mathcal{R}}\left(y_{u, t} \log \hat{y}^P_{u, t}+\left(1-y_{u, t}\right) \log \left(1-\hat{y}^P_{u, t}\right)\right),
\end{equation}
where $	\hat{y}^P_{u, t}=\sigma(\textit{logit}^P_{u, t})$ and $\mathcal{R}$ is the training set. The negative prediction tower $\hat{y}^N_{u, t}$ indeed will be self-supervised and optimized by the contrastive loss of Eq.\eqref{eq:contrastive_loss}.
After obtaining the main loss for the next item prediction, disentangling loss for repelling representations and BPR loss for pair-wise learning, we can then jointly optimize them as:
\begin{equation}\label{eq:joint_loss}
\mathcal{L}^J= \mathcal{L}+ \lambda^{BPR} \mathcal{L}^{BPR} + \lambda^{D}\mathcal{L}^D + \lambda\|\Theta\|,
\end{equation}
where $\lambda^{BPR}$ and $\lambda^{D}$ are hyper-parameters for weighting each loss. Here $\lambda$ is the regularization parameter, and $\Theta$ denotes the model parameters to be learned.

\section{Experiments}
In this section, we experiment on a public dataset and an industrial dataset, aiming to answer the following research questions (RQ): 
\begin{itemize}[leftmargin=*]
\item \textbf{RQ1}: Is the proposed DFAR effective when compared with the state-of-the-art sequential recommenders?
\item \textbf{RQ2} : What is the effect of our proposed feedback-aware encoding layer, dual-interest disentangling layer and prediction layer? 
\item \textbf{RQ3} : How do the heads of proposed feedback-aware factorization-heads attention capture the transition pattern between different feedback?
\item \textbf{RQ4}: How does the proposed method perform compared with the sequential recommenders under different sequence lengths? 
\end{itemize}
We also look into the question: "how do the auxiliary loss for disentanglement and pair-wise contrastive learning perform under different weights?" in Appendix~\ref{sec:hyper}.
\subsection{Experimental Settings}

\begin{table}[t]
\small
\caption{Micro-video and Amazon data statistics.}\label{tbl:data}
\centering
        \setlength{\belowcaptionskip}{-0.01cm}
		\setlength{\abovecaptionskip}{-0.01cm}
\begin{tabular}{cc|c|c}
\hline
\multicolumn{2}{c|}{\textbf{Dataset}}                   & \textbf{Micro-video} & \textbf{Amazon} \\ \hline
\multicolumn{2}{c|}{\textbf{\#Users}}                    & 37,497               & 6,919           \\ \hline
\multicolumn{2}{c|}{\textbf{\#Items}}                    & 129,092              & 28,695          \\ \hline
\multirow{2}{*}{\textbf{\#Records}} & \textbf{Positive} & 6,413,396            & 99,753          \\ \cline{2-4} 
                                   & \textbf{Negative} & 5,448,693            & 20,581          \\ \hline
\multicolumn{2}{c|}{\textbf{Avg.   records per user}}   & 316.35               & 17.39           \\ \hline
\end{tabular}
\end{table}
\useunder{\uline}{\ul}{}

\begin{table*}[t!]
\small
         \setlength{\belowcaptionskip}{-0.01cm}

\caption{Overall evaluations for DFAR against baselines under Micro-video and Amazon datasets {on} four metrics. Here Improv. is the improvement. Bold is the highest result and underline is the second highest result.} \label{tbl:overall}
\centering
\begin{tabular}{c|c|cccccccc|c|c}
\toprule
\multicolumn{2}{c|}{\textbf{Models}}                   & \textbf{DIN} & \textbf{Caser} & \textbf{GRU4REC} & \textbf{DIEN} & \textbf{SASRec} & \textbf{THA4Rec} & \textbf{DFN} & \textbf{FeedRec} & \textbf{Ours}   & \textbf{Improv.} \\ \midrule
\multirow{4}{*}{\textbf{Micro-video}} & \textbf{AUC}  & 0.7345       & 0.8113         & 0.7983           & 0.7446        & 0.8053          & 0.8104           & {\ul 0.8342} & 0.8119           & \textbf{0.8578} & 2.83\%            \\   
                                      & \textbf{MRR}  & 0.5876       & 0.6138         & 0.5927           & 0.5861        & 0.6046          & 0.6080           & {\ul 0.6321} & 0.6095           & \textbf{0.6568} & 3.91\%            \\   
                                      & \textbf{NDCG} & 0.6876       & 0.7079         & 0.6916           & 0.6861        & 0.7009          & 0.7035           & {\ul 0.7222} & 0.7047           & \textbf{0.7410} & 2.60\%            \\   
                                      & \textbf{GAUC} & 0.7703       & 0.8211         & 0.8041           & 0.7753        & 0.8120          & 0.8138           & {\ul 0.8362} & 0.8180           & \textbf{0.8545} & 2.19\%            \\ \midrule
\multirow{4}{*}{\textbf{Amazon}} & \textbf{AUC}  & 0.6595       & 0.7192         & {\ul 0.7278}     & 0.6688        & 0.6903          & 0.7069           & 0.6998       & 0.7037           & \textbf{0.7333} & 0.76\%            \\   
                                      & \textbf{MRR}  & 0.4344       & 0.4846         & {\ul 0.4901}     & 0.4547        & 0.4604          & 0.4599           & 0.4743       & 0.4675           & \textbf{0.4980} & 1.61\%            \\   
                                      & \textbf{NDCG} & 0.5669       & 0.6073         & {\ul 0.6114}     & 0.5832        & 0.5883          & 0.5879           & 0.5990       & 0.5938           & \textbf{0.6175} & 1.00\%            \\   
                                      & \textbf{GAUC} & 0.6618       & 0.7245         & {\ul 0.7266}     & 0.6859        & 0.7029          & 0.7021           & 0.7120       & 0.7079           & \textbf{0.7305} & 0.54\%            \\ \bottomrule
\end{tabular}
\end{table*}
\subsubsection{\textbf{Datasets}} The data statistics of our experiments are illustrated in Table~\ref{tbl:data} where Micro-video is a collected industrial dataset and Amazon is the public benchmark dataset which is widely used in existing work for sequential recommendation~\cite{lin2022dual}. The detailed descriptions of them are as below.

\para{Micro-video} This is a popular micro-video application dataset, which is recorded from September 11 to September 22, 2021. In this platform, users passively receive the recommended videos, and their feedbacks are mostly either skip or no-skip. Skip can be treated as a form of negative feedback, and no-skip can be treated as a form of positive feedback. That is to say, we have hybrid positive and negative feedback in this sequential data which is very rare in modern applications.  

\para{Amazon}\footnote{\url{https://www.amazon.com}}  This is Toys domain from a widely used public e-commerce dataset in recommendation. The rating score in Amazon ranges from 1 to 5, and we treat the rating score over three and under two as positive and negative feedback, respectively, following existing work DenoisingRec~\cite{Wang_2021} which is not for the sequential recommendation.

For the Micro-video dataset, interactions before and after 12 pm of the last day are split as the validation and test sets, respectively, while interactions before the last day are used as the training set.
For the Amazon dataset, we split the last day as the test set and the second last day as the validation set, while other days are split as the training set.

\subsubsection{\textbf{Baselines and Evaluation Metrics}}
 We compare our DFAR with the following state-of-the-art methods for sequential recommender systems. 
\begin{itemize}[leftmargin=*]
    \item \textbf{DIN}~\cite{DIN}: It aggregates the historical items via attention score with the target item.
    \item \textbf{Caser}~\cite{Caser}: It captures the transition between historical items via convolution.
    \item \textbf{GRU4REC}~\cite{GRU4REC}: It captures the transition between historical items via GRU~\cite{GRU}.
    \item \textbf{DIEN}~\cite{DIEN}: It captures the transition between historical items via interest extraction and evolution GRUs~\cite{GRU}.
    \item \textbf{SASRec}~\cite{SASRec}: It captures the transition between historical items via multi-heads attention~\cite{vaswani2017attention}.
        \item \textbf{THA4Rec}: It means talking-heads attention~\cite{shazeer2020talking} for the sequential recommendation, which is firstly applied in the recommendation by us.
    \item \textbf{DFN}~\cite{xie2021deep}: It purifies unclick (weak feedback) by click (strong positive feedback) and dislike (strong positive feedback).
    \item \textbf{FeedRec}~\cite{wu2022feedrec}: It further performs disentanglement on the weak positive and negative feedback.

\end{itemize}

Besides, Widely-used AUC and GAUC~\cite{gunawardana_evaluating_2015}  are adopted as accuracy metrics here while MRR@10 and NDCG@10~\cite{lin2022dual} are used as ranking metrics for performance evaluation. The detailed illustration of them is in Appendix~\ref{app:metric}.

\subsubsection{\textbf{Hyper-parameter Settings}} Hyper-parameters are generally set following the default settings of baselines. We strictly follow existing work for sequential recommendation~\cite{lin2022dual} and leverage Adam~\cite{Adam} with the learning rate of 0.0001 to weigh the gradients. The embedding sizes of all models are set as 32. We use batch sizes 20 and 200, respectively, on the Micro-video and Amazon datasets. We search the loss weights for pair-wise contrastive loss in $[10^{-4}, 10^{-3}, 10^{-2}, 10^{-1}]$.

\subsection{Overall Performance Comparison(RQ1)}

We compare our proposed method with eight competitive baselines, and the results are shown as Table~\ref{tbl:overall}, where we can observe that: 
\begin{itemize}[leftmargin=*]
    \item \textbf{Our method achieves the best performance.
} The results on two datasets show that our DFAR model achieves the best performance compared with these seven baselines on all metrics. Specifically, GAUC is improved by about 2.0\% on the Micro-video dataset and 0.5\% on the Amazon dataset and when comparing DFAR with other baselines. Please note that 0.5\% improvement on GAUC could be claimed as significant, widely acknowledged by existing works~\cite{DIN}. 
Besides, the improvement is more significant in the Micro-video with more negative feedback, which means incorporating the negative feedback into the historical item sequence can boost the recommendation performance.  
\item \textbf{Existing work roughly captures the relation between positive feedback and negative feedback}. FeedRec and DFN even underperform some traditional sequential recommendation models like GRU4REC and Caser in Amazon dataset. Besides, though they outperform other baselines in Micro-video dataset, the improvement is still slight. In other words, their designs fail to capture the relation between positive feedback and negative feedback, which motivates us to further improve them from transition and interest perspectives.
\end{itemize}

\subsection{Ablation Study (RQ2)}
\begin{table}[!htb]
\small
\caption{Effectiveness study of our proposed components. FHA means factorization-heads attention; MO means label mask operation on heads; IDL means interest disentangling loss on positive and negative representations; IBL means interest BPR loss on positive and negative logits.} \label{tbl:ablation}
\centering
  \setlength{\tabcolsep}{1mm}{} 
     \setlength{\belowcaptionskip}{-0.2cm}
		\setlength{\abovecaptionskip}{-0.01cm}
\begin{tabular}{c|cccc|c}
\hline
\textbf{Dataset} & \multicolumn{5}{c}{\textbf{Micro-video}}                                                   \\ \hline
\textbf{Methods} & \textbf{w/o FHA} & \textbf{w/o MO} & \textbf{w/o IDL} & \textbf{w/o IBL} & \textbf{Ours}   \\ \hline
\textbf{AUC}     & 0.8360           & 0.8473          & 0.8475           & 0.8364           & \textbf{0.8578} \\ \hline
\textbf{MRR}     & 0.6198           & 0.6378          & 0.6377           & 0.6324           & \textbf{0.6568} \\ \hline
\textbf{NDCG}    & 0.7127           & 0.7264          & 0.7264           & 0.7212           & \textbf{0.7410} \\ \hline
\textbf{GAUC}    & 0.8319           & 0.8428          & 0.8436           & 0.8283           & \textbf{0.8545} \\ \hline
\textbf{Dataset} & \multicolumn{5}{c}{\textbf{Amazon}}                                                   \\ \hline
\textbf{AUC}     & 0.7133           & 0.7141          & 0.7284           & 0.7137           & \textbf{0.7333} \\ \hline
\textbf{MRR}     & 0.4782           & 0.4883          & 0.4855           & 0.4839           & \textbf{0.4980} \\ \hline
\textbf{NDCG}    & 0.6016           & 0.6095          & 0.6073           & 0.6057           & \textbf{0.6175} \\ \hline
\textbf{GAUC}    & 0.7054           & 0.7137          & 0.7128           & 0.7047           & \textbf{0.7305} \\ \hline
\end{tabular}
\end{table}

We further study the impact of four proposed components as Table~\ref{tbl:ablation}, where FHA represents the factorization-heads attention, the MO represents the mask operation on factorized heads for factorization-heads attention, IDL means the interest disentanglement loss on the positive and negative interest representations, and IBL means the interest BPR loss on the positive and negative prediction logits. From this table, we can have the following observations.
\begin{itemize}[leftmargin=*]
     \item \textbf{Factorization of heads for transition attention weights is important}. Removing FHA and MO both show significant performance drops, which means these two components are both necessary to each other. Specifically, removing FHA means it is impossible to apply the mask on the implicit head interaction of either multi-heads attention or talking-heads attention. At the same time, removing MO on FHA will cause it to fail to exploit the prior knowledge of labels for historical items and degenerate to even as poor as multi-heads attention or talking-heads attention in the Amazon dataset.

    \item \textbf{Pair-wise interest is more important than disentangling interest}. Removing IDL and IBL will both drop the performance, while removing IBL is more significant. This is because contrastive learning by BPR loss can indeed inject more self-supervised signals, while disentanglement solely tends to repel the dissimilar representations of positive feedback and negative feedback.
\end{itemize}

\subsection{Visualization for Attention Weights of Heads (RQ3)}
\begin{figure}[!htb]
		\centering
		\begin{tabular}{cc}
		    	\includegraphics[width=0.47\columnwidth]{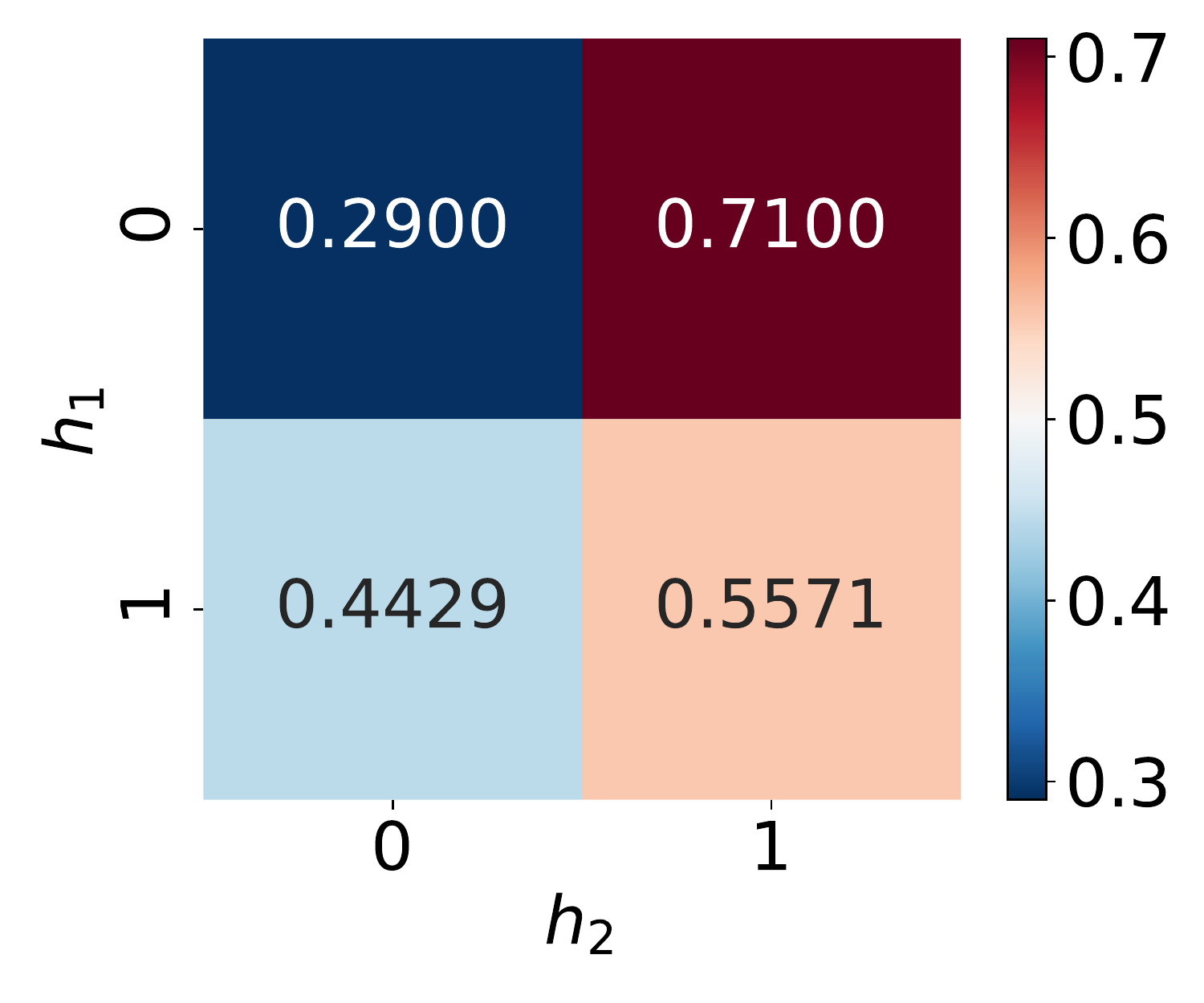} &  \includegraphics[width=0.47\columnwidth]{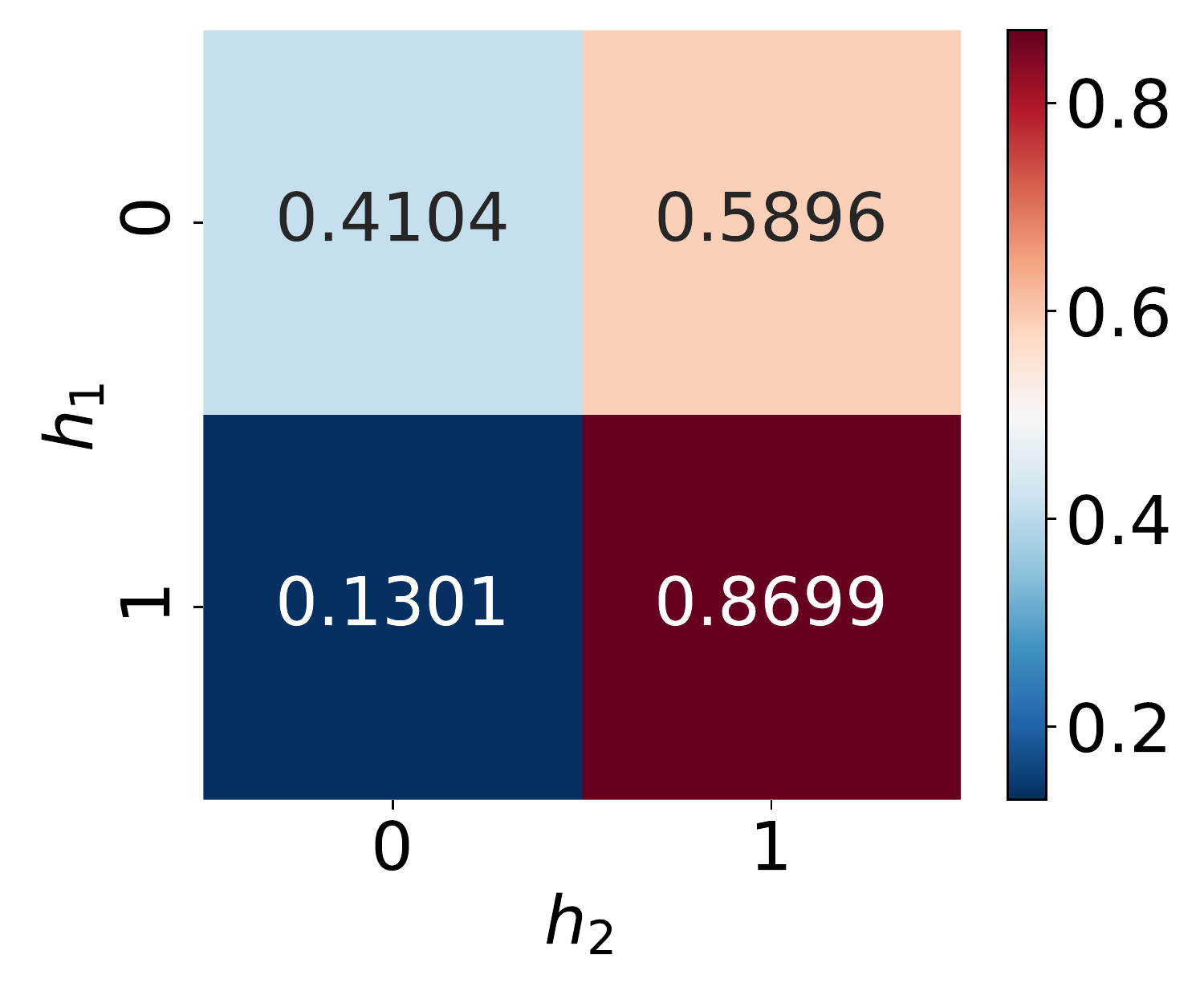} 
       \\ (a) Micro-video & (b) Amazon
		     
		\end{tabular}
		\setlength{\abovecaptionskip}{-0.01cm}
        	\caption{Visualization of accumulated attention weights between different heads. Here $h_1$ and $h_2$ represent the heads for the source and target behaviors, respectively (i.e., if the source behavior is negative and target behavior is positive, we have $h_1 = 0$ and $h_2 = 1$). This illustrates our method can factorize and extract the relation between different feedback based on the proposed factorization-heads attention.}	\label{fig:vis}
         \setlength{\belowcaptionskip}{-0.2cm}
\end{figure} 
As illustrated in Eq.\eqref{eq:fha}, our proposed factorization-heads attention can factorize the relation between different feedback, which makes it possible for us to study the attention weights between them. Therefore, we perform visualization on the attention weights between positive and negative heads in Figure~\ref{fig:vis}, where $h_1$ and $h_2$ (defined at \eqref{eq:attention_heads}) represent heads for source and target behaviors, respectively, with corresponding feedback. From this figure, we can observe that: (1) For the collected Micro-video dataset, users are still willing to watch videos even after they receive the disliked videos. This may be because the negative recommended videos are of low cost for users as they can easily skip the disliked videos, making no significant impact on their later preferred videos; (2) For the e-commerce dataset about Amazon, we can discover that when the source feedback is negative, the probability of target feedback being negative will increase sharply. This may be because the negative purchased items are of high cost in e-commerce for users as it will waste their money, increasing their unsatisfied emotion sharply. 

\subsection{The Impact of Sequence Length (RQ4)}
\begin{figure}[t]
		\centering
		\begin{tabular}{cc}
		    	\includegraphics[width=0.47\columnwidth]{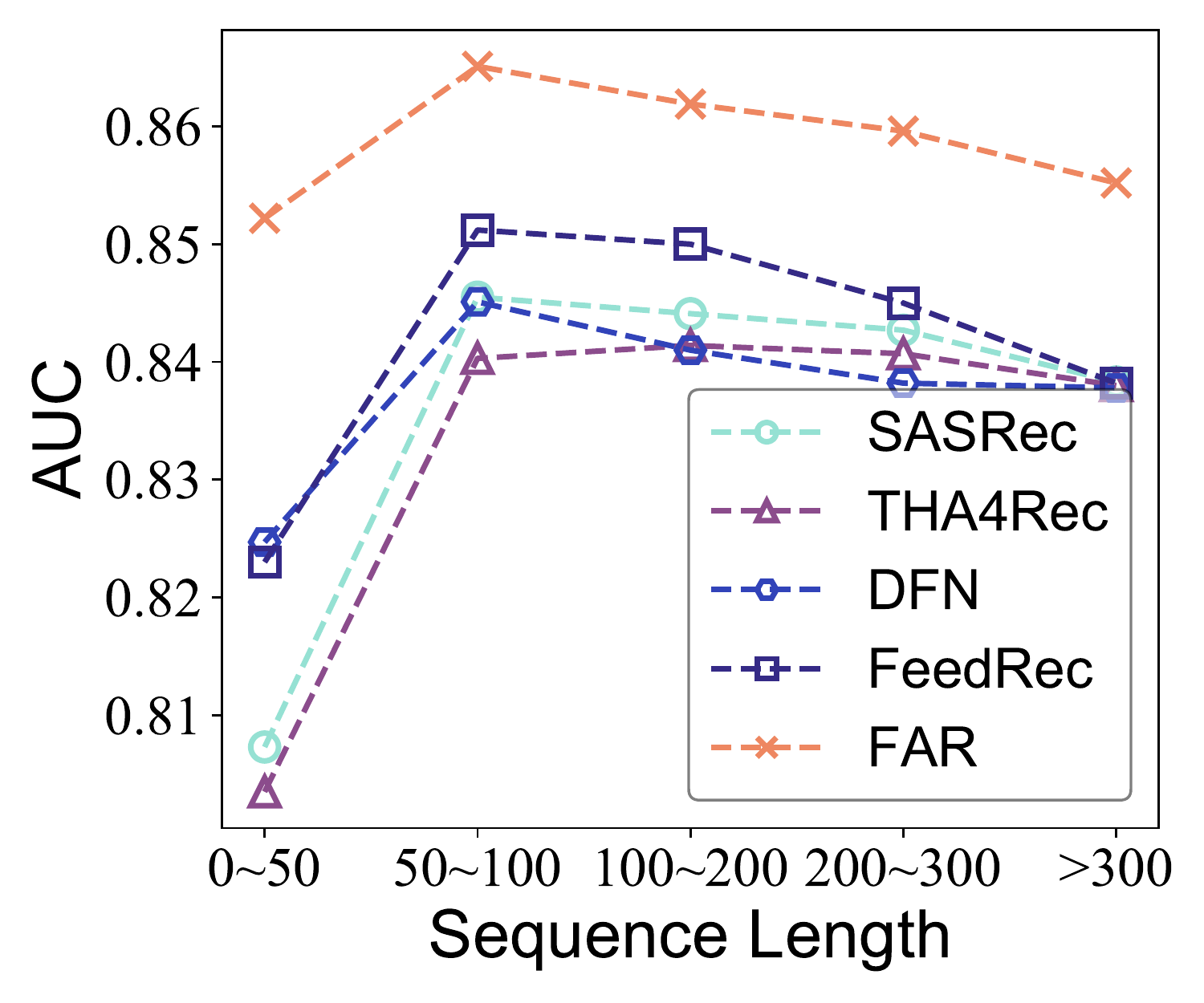} &  \includegraphics[width=0.47\columnwidth]{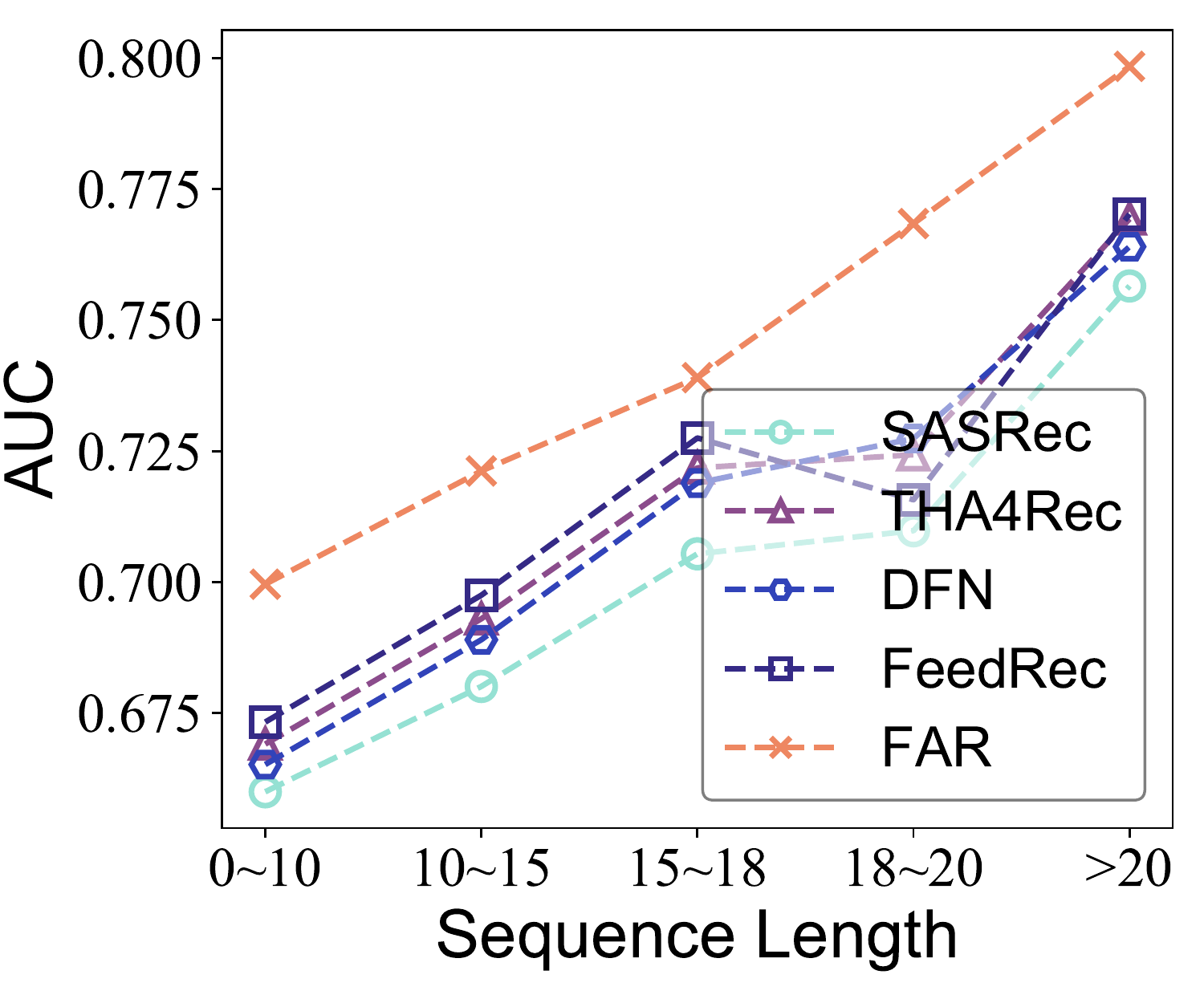} 
       \\ (a) Micro-video & (b) Amazon
		\end{tabular}
	
		\setlength{\abovecaptionskip}{-0.01cm}
        \setlength{\belowcaptionskip}{-0.2cm}
  
	\caption{AUC performance comparisons under different sequence lengths on the Micro-video and Amazon datasets.}
	      \vspace{-0.2cm}
	      \label{fig:seq}
\end{figure} 
On large-scale online platforms, active users often observe a lot of items and generate very long historical item sequences, while cold-start users are recorded with very short sequences. Long historical item sequences can bring them more information but the problem of gradient vanishing will increase, while short historical item sequence brings limited information and tends to overfit the model.
Thus, we divide historical item sequences into five groups based on their lengths and further study how DFAR outperforms the attention-based models under different lengths, under Micro-video and Amazon datasets, as illustrated in Figure~\ref{fig:seq}.
From the visualization, we can observe that:
	      \vspace{-0.05cm}
\begin{itemize}[leftmargin=*]
     \item \textbf{DFAR is superior under different sequence lengths}. It is obvious that there is always a significant performance gap between DFAR and other methods. In the Amazon dataset, where the sequence length is relatively short, the AUC performances increase with the growth of sequence length for all methods. This means a longer sequence can bring more information. However, in the Micro-video dataset where the sequence length is relatively long, the performances of all methods improve with the increase of sequence length and reach their peak at around 50-100. But then they all decline with the further increase in length. Most importantly, our DFAR outperforms other methods significantly throughout various sequence lengths.

    \item \textbf{DFAR is stable under different sequence lengths}. DFAR is more stable with the sequence length increasing or decreasing, even into very long or short. In the Amazon dataset, other methods first increase with the sequence length but fluctuate at 15-20 while DFAR increases steadily with the sequence length. In the Micro-video dataset, All methods drop sharply when the sequence length is too short or long, but our DFAR is more stable and still keeps a decent AUC performance at 0.8382.
    
\end{itemize}
	      \vspace{-0.05cm}
In summary, our DFAR is superior and robust under both long and short historical item sequences.

\vspace{-0.1cm}
\section{Related Work}
\vspace{-0.1cm}
\textbf{Sequential Recommendation}
Sequential Recommendation~\cite{SRs} predicts the next interacted item of the given user based on his/her historical items.
As the early work, FPMC~\cite{rendle2010factorizing} exploits the Markov chain to capture the transition pattern of historical item sequence in the recommendation. Then some advanced deep learning methods such as RNN~\cite{GRU, LSTM} and attentive network~\cite{vaswani2017attention} are applied in recommendation~\cite{GRU4REC, DIEN, SASRec, DIN} to capture the chronological transition patterns between historical items. While the evolution of RNN-based methods should forward each hidden state one by one and are difficult to parallel, attention-based methods can directly capture the transition patterns among all historical items at any time step.
Furthermore, researchers also attempt to leverage convolution neural network~\cite{CNN} to capture the union and point levels sequential pattern in recommendation~\cite{Caser}. Compared with CNN-based methods, attention-based methods are more effective for their non-local view of self-attention~\cite{wang2018non}.
However, the most existing sequential recommendation is based on click behavior.
Recently, there have been some methods of achieving sequential recommendations beyond click behaviors~\cite{ma2022graph}. For example, DFN~\cite{xie2021deep} captures the sequential patterns among click, unclick and dislike behaviors by an internal module for each behavior and an external module to purify noisy feedback under the guidance of precise but sparse feedback. CPRS~\cite{wu2020user} derives reading satisfaction from the completion of users on certain news to facilitate click-based modeling. Based on them, FeedRec~\cite{wu2022feedrec} further enhances sequential modeling by a heterogeneous transformer framework to capture the transition patterns between user feedback such as click, dislike, follow, etc. However, these works mainly focus on exploiting the auxiliary feedback to enhance the modeling in the sequential recommendation, which does not consider the most important characteristic - the transition patterns between historical positive and negative feedback.
Differently from them, our approach can factorize the transition patterns between different feedback, achieving more accurate modeling for sequential recommendation with both positive and negative feedback. Additionally, our approach extracts the relation between positive and negative feedback at interest level.

\noindent\textbf{Explainable Attention} Attention methods are popular in many machine learning fields such as recommender systems~\cite{SASRec, sun2019bert4rec, zhang2020general}, computer vision~\cite{fu2016aligning, wang2018non, li2019attention, dosovitskiy2021an} and natural language processing~\cite{nmt15, vinyals2015grammar}, etc. Attention mechanisms are often explainable and have been widely used in deep models to illustrate the learned representation by visualizing the distribution of attention scores or weights under specific inputs~\cite{choi2016retain, martins2016softmax, wang2016attention}.
Some explainable attention methods are also generalizable and can be equipped with many backbones. For example, L2X~\cite{chen2018learning} exploits Gumbel-softmax~\cite{jang2017categorical} for feature selection by instance, with its hard attention design~\cite{xu2015show}.
Moreover, VIBI~\cite{bang2020explaining} further propose a feature score constraint in a global prior so as to 
simplify and purify the explainable representation learning.
As self-attention is popular~\cite{vaswani2017attention, devlin-etal-2019-bert}, there is also a work that explains what heads learn and concludes that some redundant heads can be pruned~\cite{voita-etal-2019-analyzing}.
In this work, we propose feedback-aware factorization-heads attention to explicitly capture the transition pattern between positive and negative feedback. The feedback mask matrix in our attention module can be treated as hard attention based on feedback.

%\vspace{-0.1cm}

\section{Conclusions and Future Work}
\vspace{-0.1cm}
In this work, we considered the positive and negative feedback in the historical item sequence for the sequential recommendation, while existing works were mostly click-based and considered solely positive feedback. Such exploration addressed the challenge of current multi-head attention for different feedback interactions in one sequence. More specifically, we first applied talking-heads attention in the sequential recommendation and further proposed feedback-aware factorization-heads attention to explicitly achieve interaction across different heads for self-attention. Secondly, we proposed disentanglement and pair-wise contrastive learning to repel the dissimilar interests and capture the pair-wise relation between positive and negative feedback.
In the future, we plan
deploy the model in industrial applications to validate online performance.
\vspace{-0.2cm}
\section*{Acknowledgment}
\vspace{-0.1cm}
This work is supported in part by the National Key Research and Development Program of China under 2022YFB3104702, the National Natural Science Foundation of China under 62272262, 61971267, U1936217 and 61972223, BNRist, the Fellowship of China Postdoctoral Science Foundation under 2021TQ0027 and 2022M710006, and the Tsinghua University Guoqiang Institute under 2021GQG1005.

\clearpage
\appendix
\section{Appendix for Reproducibility}

\subsection{Pseudocode}\label{sec:pseudocode}

\begin{figure}[!htb]
    \small
    {
\begin{lstlisting}[
    style       =   Python,
    caption     =   {\bf Pseudocode for Multi-heads Attention},
    label       =   {code:mha},
]
def MultiHeadAttention (
X[n, d_X], # n vectors with dimensionality d_X
M[m, d_M], # m vectors with dimensionality d_M
P_q[d_X, d_k, h], # learned linear projection to produce queries
P_k[d_M, d_k, h], # learned linear projection to produce keys
P_v[d_M, d_v, h], # learned linear projection to produce values
P_o[d_Y, d_v, h]): # learned linear projection of output
    Q[n, d_k, h] = einsum (X[n, d_X], P_q[d_X, d_k, h]) 
    K[m, d_k, h] = einsum (M[m, d_M], P_k[d_M, d_k, h]) 
    V[m, d_v, h] = einsum (M[m, d_M], P_v[d_M, d_v, h]) 

    L[n, m, h] = einsum (Q[n, d_k , h], K[m, d_k , h]) # logits h*n*m* d_k

    W[n, m, h] = softmax (L[n, m, h], reduced_dim =m) # weights
    
    O[n, d_v , h] = einsum (W[n, m, h], V[m, d_v , h]) # h*n*m* d_v
    Y[n, d_Y ] = einsum (O[n, d_v , h], P_o[d_Y , d_v , h]) # output h*n* d_Y * d_v
    return Y[n, d_Y]
\end{lstlisting}}
\end{figure}
We follow talking-heads attention~\cite{shazeer2020talking} and present the following notation and pseudocode.

\subsubsection{\textbf{Notation}}
In our pseudocode, we follow talking-heads attention~\cite{shazeer2020talking} and have a notation as below.
\begin{itemize}[leftmargin=*]
    \item The capital letters represent the variable names, and lower-case letters represent the number of dimensions. Each variable of a tensor is presented with its dimensions. For example, a tensor for an item sequence with batch size $b$, sequence length $n$, hidden state $d$ is written as: X[b, n, d]~\cite{shazeer2020talking}.

    \item The einsum represents the generalized contractions between tensors without any constraint on their dimension. Its computation process is: (1) Broadcasting each input to have the union of all dimensions, (2) multiplying component-wise, and (3) summing across all dimensions not in the output. The dimensions are identified by the dimension-list annotations on the arguments and on the result instead of being identified by an equation, as in TensorFlow and NumPy. For example, multiplying two matrices is written as: Z[a, c] = einsum (X[a, b], W[b, c])~\cite{shazeer2020talking}.

\end{itemize}

\subsubsection{\textbf{Multi-heads Attention}}

The pseudocode for multi-heads attention~\cite{vaswani2017attention} is as shown in Pseudocode~\ref{code:mha}, where different heads for Q and K do not interact with each other on line 12.

\subsubsection{\textbf{Talking-heads Attention}}

The pseudocode for talking-heads attention~\cite{shazeer2020talking} is as shown in Pseudocode~\ref{code:tha}, where different heads for Q and K achieve implicit interaction by lines 15 and 18.

\subsubsection{\textbf{Factorization-heads Attention}}

The pseudocode for our proposed factorization-heads attention is as shown in Pseudocode~\ref{code:fha}, where different heads for Q and K achieve explicit interaction by line 16. 
\begin{figure}[!htb]
    \small
    {
\begin{lstlisting}[
    style       =   Python,
    caption     =   {\bf Pseudocode for Talking-heads Attention},
    label       =   {code:tha},
]
def TalkingHeadAttention (
X[n, d_X], # n vectors with dimensionality d_X
M[m, d_M], # m vectors with dimensionality d_M
P_q[d_X, d_k, h_k], # learned linear projection to produce queries
P_k[d_M, d_k, h_k], # learned linear projection to produce keys
P_v[d_M, d_v, h_v], # learned linear projection to produce values
P_o[d_Y, d_v, h_v]
P_l [h_k , h], # talking - heads projection for logits
P_w [h, h_v]): # talking - heads projection for weights
    Q[n, d_k, h_k] = einsum (X[n, d_X], P_q[d_X, d_k, h_k]) 
    K[m, d_k, h_k] = einsum (M[m, d_M], P_k[d_M, d_k, h_k]) 
    V[m, d_v, h_v] = einsum (M[m, d_M], P_v[d_M, d_v, h_v]) 

    J[n, m, h_k] = einsum (Q[n, d_k, h_k], K[m, d_k, h_k]) # dot prod . n*m* d_k *h_k
    L[n, m, h] = einsum (J[n, m, h_k], P_l [h_k, h]) # Talking - heads proj . n*m*h* h_k

    W[n, m, h] = softmax (L[n, m, h], reduced_dim=m) # Attention weights
    U[n, m, h_v] = einsum (W[n, m, h], P_w [h, h_v]) # Talking - heads proj . n*m*h* h_v
    
    O[n, d_v, h_v] = einsum (U[n, m, h_v], V[m, d_v, h_v]) # n*m* d_v * h_v
    Y[n, d_Y] = einsum (O[n, d_v, h_v], P_o [d_Y, d_v, h_v]) # n* d_Y * d_v * h_v
    return Y[n, d_Y]
\end{lstlisting}}
\end{figure}

\subsubsection{\textbf{Comparison}}\label{sec:compare}
From these three Python pseudocodes, we can discover that our factorization-heads attention achieves head interaction at a low cost. The comparison of it with multi-heads attention and talking-heads attention are as below.
\begin{itemize}[leftmargin=*]
    \item \textbf{Comparing with Multi-heads Attention}: our factorization-heads attention incorporates the interaction between different heads with additional four lines at lines 12-14 and 17, which are transpose and reshape operations and with only $O(1)$ temporal complexity.
~\footnote{\url{https://stackoverflow.com/questions/58279082/time-complexity-of-numpy-transpose} }.
\item \textbf{Comparing with Talking-heads Attention}: our factorization-heads attention achieves explicit interaction with additional transpose and reshape operations at $O(1)$ temporal complexity while talking-heads attention achieves implicit interaction with two matrix multiplication operations at $O(m \times h_k \times h)$ and $O(m \times h \times h_v)$ temporal complexities~\footnote{\url{https://en.wikipedia.org/wiki/Computational_complexity_of_matrix_multiplication}}, respectively.

\end{itemize}
\subsection{Evaluation Metrics}\label{app:metric}

The detailed illustration of adopted evaluation metrics is as follows.
\begin{itemize}[leftmargin=*]
     \item \textbf{AUC}: Randomly selecting one positive item and one negative item, it represents the probability that the predicted score of the positive item is higher than that of the negative item. It tests the model's ability to classify the positive and negative items.
     \item \textbf{GAUC}: It weighs each user's AUC based on his/her test set size. It tests the model's personalized classification ability on each user as recommender systems indeed tend to rank preferred items for users individually.
     \item \textbf{MRR@K}: It is the average of the reciprocal of the first hit item ranking.
     \item \textbf{NDCG@K}: It assigns hit items that rank higher with more weights and thus tests the model's ability to rank the hit items in higher and more confident positions. 
 \end{itemize}

\subsection{Implementation Details}
We implement all the models by a Microsoft~\footnote{\url{https://github.com/microsoft/recommenders}} TensorFlow~\footnote{\url{https://www.tensorflow.org}} framework in Python, which is accessible here~\footnote{\url{https://anonymous.4open.science/r/DFAR-8B7B}}. We will publish the Micro-video dataset to benefit the community in the future, and the public Amazon dataset is accessible at this website~\footnote{\url{http://jmcauley.ucsd.edu/data/amazon/index_2014.html}}.

The environment is as below.
\begin{itemize}
    \item Anaconda 3
\item Python 3.7.7
\item TensorFlow 1.15.0
\end{itemize}

Besides, for other parameters, we stop the model training with early stop step 2 and leverage the MLP layer sandwiched between two normalization layers as the prediction tower for each model.

\subsection{Hyper-parameter Study (RQ5)} \label{sec:hyper}
\begin{figure}[!htb]
		\centering
		\begin{tabular}{cc}
		    	\includegraphics[width=0.47\columnwidth]{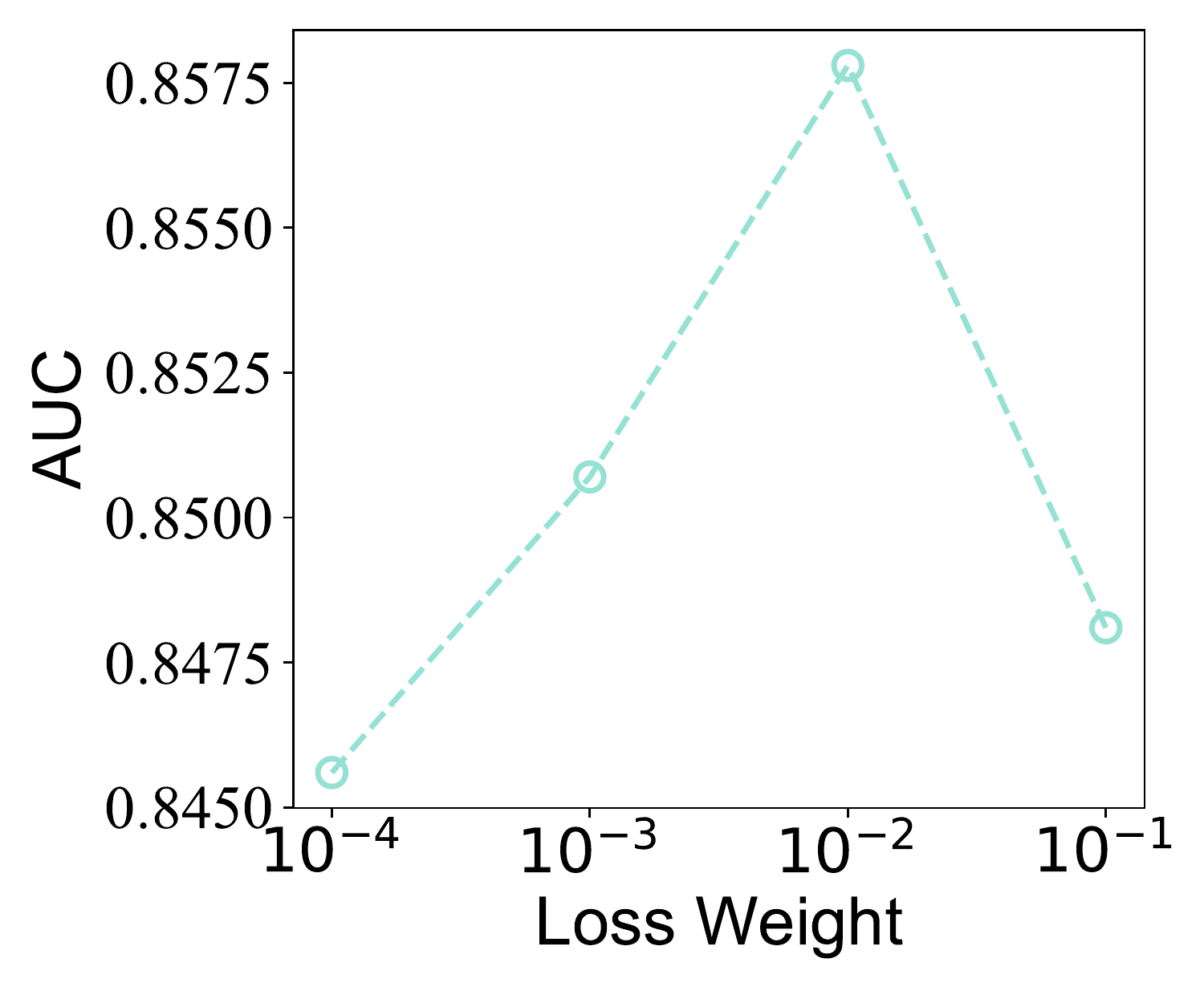} &  \includegraphics[width=0.47\columnwidth]{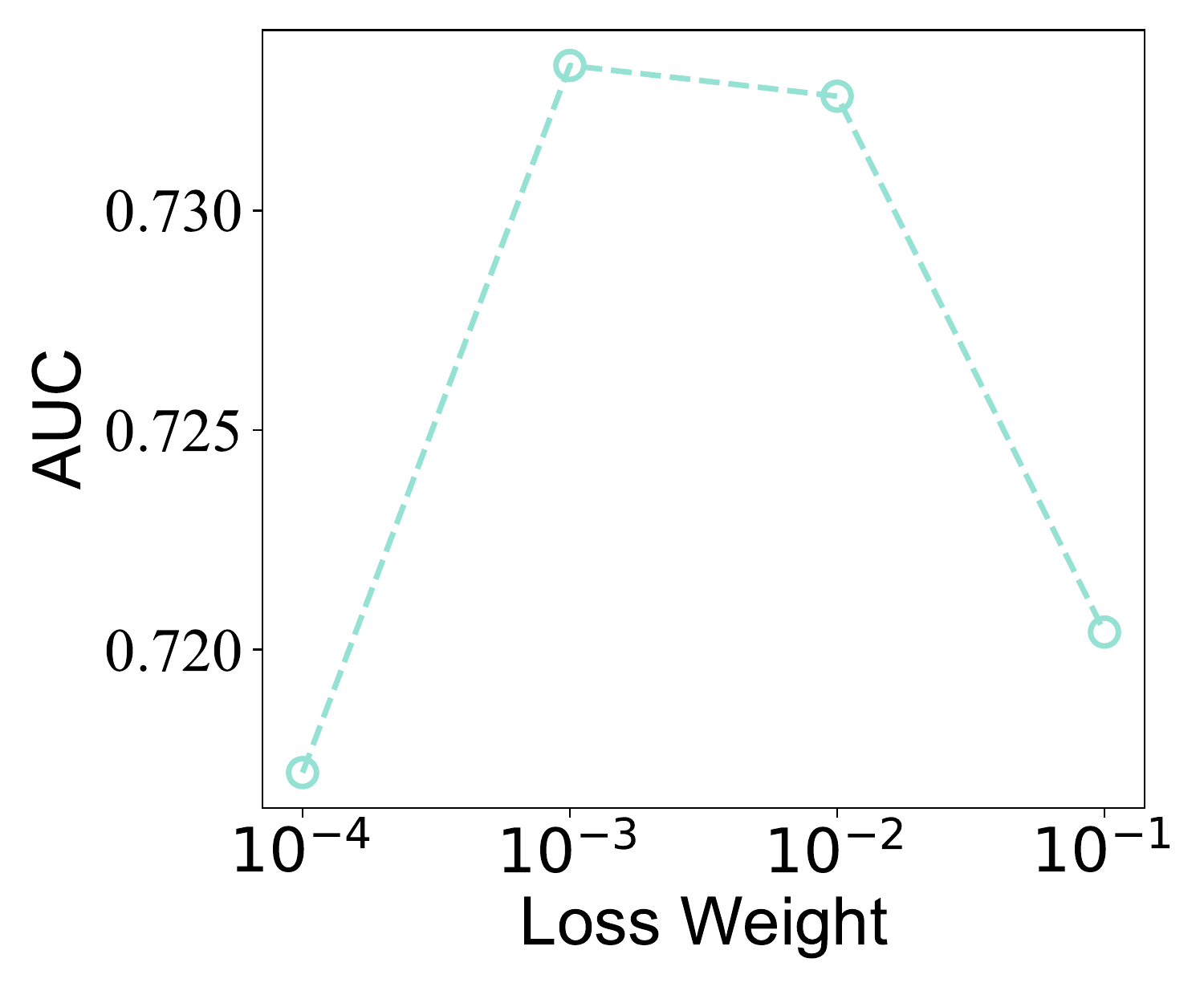} 
		     \\ (a) Micro-video & (b) Amazon
		\end{tabular}

	\caption{AUC performance of different auxiliary loss weights w.r.t $\lambda^{BPR}$ and $\lambda^{D}$ under Micro-video and Amazon datasets.}	\label{fig:hyper}
\end{figure} 

We perform hyper-parameter study on the weights for loss of disentanglement and pair-wise contrastive Learning (w.r.t. $\lambda^{BPR}$ and $\lambda^{D}$ at Eq.\eqref{eq:joint_loss}) as Figure~\ref{fig:hyper}, varying the loss weights from $10^{-4}$ to $10^{-1}$. From the figure, we can observe that the AUC performance reaches the peak at $10^{-3}$ under the Amazon dataset while that reaches the peak at $10^{-2}$ under the Micro-video dataset. 
This is because the rating for Amazon is a discrete value, but the playing time for Micro-video is a continuous value. The partition of positive and negative feedback based on continuous value is unclear and thus requires more contrastive learning.
Based on the above observation, we finally choose $10^{-3}$ and $10^{-2}$ as the best values for the loss weights under Amazon and Micro-video datasets, respectively.

\begin{figure}[!htb]
    \small
    {
\begin{lstlisting}[
    style       =   Python,
    caption     =   {\bf Pseudocode for Factorization-heads Attention},
    label       =   {code:fha},
]
def FactorizationHeadAttention (
X[n, d_X], # n vectors with dimensionality d_X
M[m, d_M], # m vectors with dimensionality d_M
P_q[d_X, d_k, h], # learned linear projection to produce queries
P_k[d_M, d_k, h], # learned linear projection to produce keys
P_v[d_M, d_v, h], # learned linear projection to produce values
P_o[d_Y, d_v, h]): # learned linear projection of output
    Q[n, d_k, h] = einsum (X[n, d_X], P_q[d_X, d_k, h]) 
    K[m, d_k, h] = einsum (M[m, d_M], P_k[d_M, d_k, h]) 
    V[m, d_v, h] = einsum (M[m, d_M], P_v[d_M, d_v, h]) 

    Q[n, h, d_k] = reshape(transpose(Q, [0, 2, 1]), [n * h, d_k]) # queries h*n* d_X * d_k
    K[d_k, h, m] = reshape(transpose(K, [1, 2, 0]), [d_k, h * m]) # keys h*m* d_M * d_k
    V[m, d_v, h * h] = tile(V[m, d_v, h], [1, 1, h]) # values h*m* d_M * d_v

    L[n * h, h * m] = einsum (Q[n * h, d_k], K_[d_k, h * m]) 
    L[n, h * h, m] = transpose(reshape(L, [n, h * h, m]), [0, 2, 1]) # logits h*h*n*m* d_k

    W[n, m, h * h] = softmax (L[n, m, h * h], reduced_dim=m) # weights
    O[n, d_v , h * h] = einsum (W[n, m, h * h], V[m, d_v, h * h]) # h*h*n*m* d_v
    Y[n, d_Y] = einsum (O[n, d_v, h * h], P_o[d_Y, d_v, h * h]) # output h*h*n* d_Y * d_v
    return Y[n, d_Y]
\end{lstlisting}}
\end{figure}

\clearpage
\balance
\bibliographystyle{ACM-Reference-Format}
\bibliography{sample-base}
\clearpage

\end{document}